\documentclass[10pt,aps,prd,superscriptaddress,nofootinbib,nobibnotes,longbibliography,floatfix,twocolumn]{revtex4-2}

\usepackage{bm}
\usepackage{mathtools,
amssymb,
amsfonts,
mathrsfs,
chngcntr,
multirow}

\let\cc\corresponds
\let\corresponds\relax
\usepackage{mathabx}
\let\corresponds\cc

\usepackage[utf8]{inputenc}
\usepackage[T1]{fontenc}

\usepackage{soul}

\usepackage[dvipsnames]{xcolor}
\usepackage[unicode]{hyperref}
\hypersetup{colorlinks=true, citecolor=MidnightBlue,
            linkcolor=MidnightBlue, urlcolor=MidnightBlue, linktocpage=true}
\usepackage[normalem]{ulem}

\usepackage{graphicx}

\usepackage{amsthm}

\usepackage{enumitem}

\newcommand{\PRLsection}[1]{\noindent{\bf\emph{#1.---}}}

\begin{document}
\title{Higher-derivative gravitational effective field theories are generically weakly hyperbolic}

\author{Farid Thaalba}
\affiliation{SISSA, Via Bonomea 265, 34136 Trieste, Italy and INFN Sezione di Trieste}
\affiliation{Nottingham Centre of Gravity \& School of Mathematical Sciences, University of Nottingham, University Park, Nottingham NG7 2RD, United Kingdom}

\author{Fernando Abalos}
\affiliation{Departament de F\'isica, Universitat de les Illes Balears, Palma de Mallorca, E-07122, Spain and Institute of Applied Computing and Community Code (IAC3), Universitat de les Illes Balears, Palma de Mallorca, E-07122, Spain} 

\author{Miguel~Bezares}
\affiliation{Nottingham Centre of Gravity \& School of Mathematical Sciences, University of Nottingham, University Park, Nottingham NG7 2RD, United Kingdom}
\begin{abstract}
We analyse the initial-value problem of metric higher-derivative effective theories of gravity. We show that any such theory whose characteristic velocities are independent of derivatives of the metric is intrinsically weakly hyperbolic, independently of the gauge fixing. To show this, we identify the spin-$2$ physical sector directly from the characteristic equation; this can be done without introducing an order-reduced formulation, which greatly simplifies the computation. In this sector, every metric theory with more than two derivatives in the equations of motion contains a weakly hyperbolic block. Since this obstruction is physical, no choice of gauge or constraint addition can remove it, providing a structural explanation for the failure of strong hyperbolicity in this broad class of theories.  
\end{abstract}
\maketitle
%-------------------------------------------------------------------------------------------------
%-------------------------------------------------------------------------------------------------
\PRLsection{Introduction} At present, within the general theory of relativity, numerical simulations of binary black hole (BBH) mergers are routinely performed with great success~\cite{Scheel:2025jct}. Notwithstanding, it took considerable effort to devise appropriate formulations and numerical techniques capable of such simulations~\cite{Pretorius:2005gq,Campanelli:2005dd, Baker:2005vv}. Numerical relativity~\cite{Alcubierre:1138167,bona2009elements,shibatabook,Baumgarte:2010ndz,Palenzuela:2020tga, Hilditch:2024nhf} is invaluable in enabling precise simulations of scenarios that are otherwise intractable, such as gravitational collapse~\cite{Choptuik:1992jv, Choptuik:2003ac,Gundlach:2025yje} and compact object mergers~\cite{ Centrella:2010mx, Lehner:2014asa, Bambi:2025btf}. Central to this progress and to achieving this milestone are the seminal works of~\cite{Foures-Bruhat:1952grw, Choquet-Bruhat:1969ywq, Sbierski:2013kca} in formulating a well-posed initial value problem (IVP) in general relativity (GR), see~\cite{Rendall:2002yr, Sarbach:2012pr, Isenberg:2013iva} for a review. 

A nonnegotiable property of any meaningful theory describing a physical system is its ability to make predictions given enough information. Formally, this is precisely reflected in the well-posedness of the IVP with suitable initial data. The notion of a (locally) well-posed IVP corresponding to a system of partial differential equations (PDEs) dates back to Hadamard~\cite{hadamard}, who required the PDE system to be uniquely solvable and continuously dependent on the initial data (given a specific function space or a norm).

As of late, spurred by the detection of gravitational waves~\cite{KAGRA:2021vkt, LIGOScientific:2018mvr, LIGOScientific:2020ibl, LIGOScientific:2021usb}, tremendous effort has been poured into testing GR~\cite{Berti:2015itd, Endlich:2017tqa, Barack:2018yly, LIGOScientific:2021sio, LISA:2022kgy, LISA:2024hlh, Yunes:2024lzm, ET:2025xjr,Barausse:2020rsu}. Most current tests of GR assume its predictions as the null hypothesis. Nonetheless, obtaining waveforms beyond GR permits more accurate and precise tests. Many studies explored the well-posedness of the IVP in various extenstions of GR~\cite{Capozziello:2011et, Burgess:2003jk, Kobayashi:2019hrl, Padmanabhan:2013xyr, Sotiriou:2008rp, Clifton:2011jh,Yunes:2024lzm}; some are numerical~\cite{Barausse:2012da,Palenzuela:2013hsa,Ripley:2019hxt, Ripley:2022cdh, Bezares:2020wkn, Bernard:2019fjb, AresteSalo:2022hua, R:2022hlf, Cayuso:2017iqc, Cayuso:2020lca, Franchini:2022ukz, Cayuso:2023xbc, Lara:2021piy, Thaalba:2023fmq, Thaalba:2024crk, Gerhardinger:2022bcw, deRham:2023ngf, Doneva:2023oww, Corelli:2025kms,Shum:2025lgp}, while others are more analytical~\cite{Papallo:2017qvl, Papallo:2017ddx, Kovacs:2020ywu, Kozuszek:2024vyb, LeFloch:2014zva, LeFloch:2025sih, Luz:2025rqx, Figueras:2024bba}.

Most extensions of GR are seen through the prism of effective field theories (EFTs)~\cite{Burgess:2007pt, Donoghue:1994dn, Burgess:2003jk}. EFTs provide controlled
parameterizations of low-energy deviations from GR. This agnostic stance toward the ultraviolet completion of gravity enables us to constrain such deviations empirically, using a finite set of parameters~\cite{Burgess:2007pt}. 

In vacuum, the leading-order corrections to GR are quadratic in curvature, leading to higher-order PDEs~\cite{Donoghue:1994dn, Burgess:2003jk, Stelle:1976gc, Noakes:1983xd}. On the other hand, in some cases, including specific (non-minimal) coupling with additional fields, e.g., a scalar field, produces second-order PDEs~\cite{Horndeski:1974wa, Weinberg:2008hq, Deffayet:2009mn};
nevertheless, in such theories, the characteristic velocities (i.e., the speeds of high-frequency perturbations) generally depend on field derivatives~\cite{Reall:2021voz, Tanahashi:2017kgn}; hence, they are not necessarily well-posed. Thus, in broad terms, we may distinguish two scenarios: either the characteristics depend on derivatives of the metric or other fields, potentially leading to well- or ill-posed IVPs and to shock formation~\cite{Reall:2014sla,Tanahashi:2017kgn}; or they do not. 

The principal symbol (i.e., the highest derivative terms) of a PDE defines its hyperbolic character~\cite{Kreiss, Sarbach:2012pr, Hilditch:2013sba}. In general, the equations of an IVP are either strongly or weakly hyperbolic. Roughly speaking, a weakly hyperbolic system might, under additional conditions on the lower-order terms, admit a well-posed IVP. In contrast, a strongly hyperbolic PDE system implies a well-posed IVP regardless of the lower-order terms (more details in the main text).

To study the hyperbolicity of second-order (or higher) PDE systems, it is usually necessary to reduce the system to first-order (at least in time)~\cite{Kreiss, Sarbach:2012pr, Hilditch:2013sba, Abalos:2017jtt}. However, by employing the ``Matrix Pencils'' technique~\cite{Abalos:2018uwg, Abalos:2021rqs, Abalos:2024auy, Abalos:2024auy, Abalos:2026kim}, one can avoid this complication and provide an equivalent notion of weak and strong hyperbolicity~\cite{Abalos:2026kim}.\footnote{Here, equivalence refers to a first-order reduction in time and space.}

Another central subtlety in formulating the IVP of any gauge-invariant theory is the separation of physical degrees of freedom from gauge redundancy. This issue is especially manifest in GR and its extensions, where different formulations and gauges might alter the character of the equations and affect the well-posedness of the IVP~\cite{Pretorius:2005gq,Baumgarte:1998te,Nakamura:1987zz,Alic:2011gg, christodoulou2008mathematical, Papallo:2017qvl, Kovacs:2020ywu, Reall:2021voz, Thaalba:2023fmq, Thaalba:2024crk, Doneva:2024ntw}.

Here, we separate the spin-$2$ physical characteristic speeds from gauge and constraints effects.\footnote{Henceforth, we only study spin-$2$ modes. The spin-$0$ modes do not share the subtleties of the spin-$2$ modes discussed below.} We then show that purely metric vacuum higher-derivative gravitational EFTs, whose characteristic speeds do not depend on metric derivatives, are weakly hyperbolic in \textit{any} gauge.

%
%
%
%-------------------------------------------------------------------------------------------------
%-------------------------------------------------------------------------------------------------
\PRLsection{Preliminaries}
In this work, we focus on purely metric gravitational EFTs with an arbitrary number of derivatives. If the original equations are of order \(q\), we collect the metric and its derivatives up to order \(q-1\) into a vector \(\mathbf u\). The equations can then be written as a first-order system
\begin{align}
    \label{eq:linear_sys_1}
    \partial_t \mathbf{u}
    +
    \mathbf{P}^{j}\partial_j \mathbf{u}
    =
    \mathbf{f},
\end{align}
where \(\mathbf u\) is an \(N\)-component vector, \(\mathbf P^j\) are \(N \times N \) constant real matrices and $\mathbf{f}$ is an $N$ source vector with no derivatives of $\mathbf{u}$.  The associated IVP is well-posed if the solution satisfies
\begin{align}
    \label{eq:estimate}
   E(t)
\leq
C e^{\alpha t}
\left[
E(0)
+
\int_0^t
\|\mathbf f(\tau,\cdot)\|_{L^2}^2
\,d\tau
\right],
\end{align}
with $E(t)\equiv \|\mathbf{u}(t,\cdot)\|_{L^2}^2$ and \(C,\alpha>0\) are constants independent of the initial data. Written in terms of the original variables, this \(L^2\) estimate becomes a Sobolev estimate. Thus, for example, in a fourth-order evolution equation, the variable $\mathbf u$ includes  $
    \sim
    \left(
    u,
    \partial u,
    \partial^2 u,
    \partial^3 u
    \right),$
and the basic (third-order Sobolev) energy corresponds to
\begin{align}
    \label{eq:norm}
    E(t)_{\mathcal{H}^3}
    &\equiv
    \|u(t, \cdot)\|_{H^3}^2
    +\|\partial_t u(t, \cdot)\|_{H^2}^2 \nonumber \\
    &+\|\partial_t^2u(t, \cdot)\|_{H^1}^2 +\|\partial_t^3u(t, \cdot)\|_{L^2}^2~,
\end{align}
which is the minimum requirement to ensure that all the derivatives of the initial data, i.e., at $t=0$ (up to third-order), are bounded in $L_2$. Here \(H^m(\Sigma_t)\) denotes the standard Sobolev space on \(\Sigma_t\).\footnote{The Sobolev norm \(H^m(\Sigma_t)\) is defined by
\[
\|v\|_{H^m(\Sigma_t)}^2
=
\sum_{|\alpha|\leq m}
\|\partial_x^\alpha v\|_{L^2(\Sigma_t)}^2 .
\]}
For system \eqref{eq:linear_sys_1}, the well-posedness is studied in the high-frequency limit and is captured by some algebraic properties of the principal symbol $P(k)=\mathbf P^j k_j$, where $k_j$ is the real spatial wave covector. Weak hyperbolicity requires the eigenvalues of \(P(k)\), which define the characteristic speeds, to be real for every real spatial covector \(k_j\). Strong hyperbolicity requires real characteristic speeds and a complete set of characteristic modes, or equivalently, a uniform diagonalizability of $P(k)$ for every real $k_j$. Alternatively, the condition of strong hyperbolicity becomes that the eigenvalues of the second-order matrix pencil (principal symbol) have equal algebraic and geometric multiplicities~\cite{Abalos:2026kim}. This condition guarantees the energy estimate~\eqref{eq:estimate}, and hence well-posedness in $L^2$ for \textit{arbitrary source terms}  (or lower-order terms) $\mathbf f$~\cite{Kreiss,Sarbach:2012pr,Hilditch:2013sba}, whereas weak hyperbolicity alone does not.

However, weakly hyperbolic systems may, in special cases and for restricted choices of the lower-order terms, satisfy estimates with a finite loss of derivatives, e.g., controlling \(\|u(t)\|_{L^2}\) by \(\|u(0)\|_{H^m}\) for some \(m>0\). Such estimates are weaker than standard Sobolev well-posedness and are not stable under arbitrary lower-order perturbations~\cite{Kreiss}. In the following, for an $n$-th order PDE system, we only consider the $(n-1)$ Sobolev energy $E(t)_{\mathcal{H}^{n-1}}$, i.e., the trivial generalization of~\eqref{eq:norm}.

As we focus on EFTs of gravity, we consider the following diffeomorphism-invariant action for a gravitational theory of only the metric $g_{ab}$
\begin{align}
    \label{eq:general_L}
     S = \int \mathrm{d}^4 x \sqrt{\vert g\vert } \mathcal{L}(g)~,
\end{align}
with its field equations given by
\begin{equation}
\label{Eq_sys_1}
    E^{ab} \equiv \frac{\delta S}{\delta g_{ab}} = \mathfrak{N}^{abc_{1} \cdots c_{n}ef}\partial_{c_{1}} \cdots \partial
    _{c_{n}}g_{ef}+ \cdots=0~. %
\end{equation}
As discussed in appendix~\ref{sec:syms}, (some of) the symmetries of $\mathfrak{N}^{abc_{1} \cdots c_{n}ef}$ are
\begin{align}
0  &  =\mathfrak{N}^{a(bc_{1} \cdots c_{n})ef}~,\label{id_bia_1}\\
0  &  =\mathfrak{N}^{ab(c_{1} \cdots c_{n}e)f}~. \label{id_gau_1}
\end{align}
We define a local $3+1$ splitting of spacetime such that it is foliated by a collection of spacelike hypersurfaces $\Sigma_{t}$~\cite{gourgoulhon2012}. The normal to these hypersurfaces is denoted $n_{a}$ with norm $n_{a}n^{a}=-1$, and the projector is $\gamma
_{ab}=g_{ab}+n_{a}n_{b}$ over $\Sigma_{t}$. We assume that
\begin{equation}
    \gamma_{c}^{a}\gamma_{d}^{b}E^{cd}=0~, \label{Physical_eq_1}%
\end{equation}
are $n$th-order evolution PDEs, that is, if we define $\partial_{0} \equiv~n^{a}\partial_{a}$, then these equations include $n$
derivatives in the $n^{a}$ direction for the six components of $\gamma_{ij}$, i.e., we have $\partial^{n}_{0}\gamma_{ij} \equiv n^{c_{1}}\cdots n^{c_{n}}\partial_{c_{1}} \cdots \partial_{c_{n}}\gamma_{ij}$.
To solve these equations, we specify $\left(  \gamma_{ij},\partial_{0}\gamma_{ij}, \cdots,\partial_{0}^{n-1}\gamma_{ij}\right)$ as initial data\footnote{Additional data must be provided after gauge fixing.}. We will discuss the hyperbolicity of~\eqref{Physical_eq_1} without choosing a particular gauge. Contracting~\eqref{Eq_sys_1} with $n_a$ defines the constraints
\begin{align}
    C^{b}\equiv n_{a}\mathfrak{N}^{abc_{1} \cdots c_{n}ef}\partial_{c_{1}} \cdots \partial_{c_{n}}g_{ef}+ \cdots=0~,
\end{align}
since $C^{b}$ do not include terms with $\partial^{n}_{0}$, they are not evolution equations. Terms with $\partial^{n}_{0}$ disappear due to the symmetries~\eqref{id_bia_1}, as explained by Geroch~\cite{Geroch:1996kg}. Thus, we need to choose $\left(  \gamma_{ij},\partial_{0}\gamma_{ij},...,\partial_{0}%
^{n-1}\gamma_{ij}\right)  $ over $\Sigma_{0}$ such that $C^{b}=0$, and due to the generalized Bianchi identities (see appendix~\ref{sec:syms}), the constraints are preserved over $\Sigma_{t},$ for $t>0$. 

We consider now the characteristic equation associated with~\eqref{Eq_sys_1}, that is
\begin{equation}
\label{eq_char_1}
    \mathfrak{N}^{abc_{1} \cdots c_{n}ef}l_{c_{1}} \cdots l_{c_{n}}\delta g_{ef}
    =0~,
\end{equation}
where $\mathfrak{N}^{abc_{1} \cdots c_{n}ef}$ is evaluated on a spacetime
background solution $g^{\text{BG}}_{ab}$, with a high frequency perturbation $\delta g_{ef}$, for
\begin{align}
    l_{a}\equiv\lambda n_{a}+k_{a}~, \label{eq_l}
\end{align}
where $\lambda \in \mathbb{R}$ is a parameter and $k_{a}$ is spacelike covector, which is orthonormal to
$n_{a}$ (with respect to $g^{\text{BG}}_{ab}$). The matrix $\mathfrak{N}^{abc_{1} \cdots c_{n}ef}l_{c_{1}} \cdots l_{c_{n}}$ is an $n$-th order polynomial in $\lambda$, the so-called $n$-th order pencil~\cite{Abalos:2026kim}. Solving equation~\eqref{eq_char_1}
amounts to finding $\lambda$ and $\delta g_{ef}$. To study the possible solutions of equation~\eqref{eq_char_1}, we define an
orthonormal basis of $T_{p}^{\ast}M$ as $\left\{n_{a},~k_{a},~e_{1a},~e_{2a}\right\}$ where $e_{1,2}$ are spacelike covectors. Using identities~\eqref{id_bia_1} and~\eqref{id_gau_1}, we have
\begin{align}
    \mathfrak{N}^{abc_{1} \cdots c_{n}ef}l_{c_{1}} \cdots l_{c_{n}}l_{e}= 0 = l_{b}%
    \mathfrak{N}^{abc_{1} \cdots c_{n}ef}l_{c_{1}} \cdots l_{c_{n}}~, \label{eq_gau_bin_1}
\end{align}
thus, we only need to consider the subspace orthogonal to the vector $l$ when solving the characteristic equation~\eqref{eq_char_1}.

%-------------------------------------------------------------------------------------------------
%-------------------------------------------------------------------------------------------------
\PRLsection{Physical sector}
The preservation of the Hamiltonian constraint \(C_H~\equiv~n_a n_b E^{ab}\) is guaranteed once the momentum constraints \(C_M^{a}~\equiv~n_c \gamma^a{}_{b} E^{cb}\) are preserved. This follows from the generalized Bianchi identity, which implies $0~=~-~n^a\nabla_a C_H~+~\nabla_b C_M^b~+~\cdots$, where, on-shell, the ellipsis denotes terms homogeneous in the constraints. This constraint-propagation equation might be modified by constraint addition, namely by adding suitable multiples of the constraints to the evolution equations; see~\cite{Abalos:2021rqs} for more details.
Let us now assume that equations~\eqref{Physical_eq_1} form a system of six \(n\)-th order evolution PDEs for the six independent components of \(\gamma_{ij}\). At the linearized level, three of these equations propagate the momentum constraints \(C_M^{a}\), and hence also the Hamiltonian constraint, as explained above. The remaining three equations decompose into two physical equations, whose characteristic structure is independent of constraint addition, and one auxiliary equation. The characteristic structure of the auxiliary equation, together with the remaining three constraint-preserving evolution equations, can be modified through constraint addition. 

Therefore, the physical velocities remain unchanged after the gauge is fixed and under constraint addition. We will use this idea to identify the $2 \times 2 $ physical block and its characteristic velocities.\footnote{After gauge fixing, the bases to obtain the physical block may change; however, they always exist for an appropriately chosen gauge.} To that end, note first that any physical mode~$\delta g_{ef}$ in~\eqref{eq_char_1} should satisfy the constraint characteristic equations for any value of $\lambda$, i.e.,
\begin{align}
\label{eq:char_const}
n_{b}\mathfrak{N}^{abc_{1} \cdots c_{n}def}l_{c_{1}} \cdots l_{c_{n}}\delta g_{ef}=0~.
\end{align}
From equation~\eqref{eq_gau_bin_1} and~\eqref{eq_l}, we have
\begin{align}
    0&=n_{a}l_{b}\mathfrak{N}^{abc_{1} \cdots c_{n}ef}l_{c_{1}} \cdots l_{c_{n}} \nonumber \\
    &=n_{a}\left(\lambda n_{b} + k_{b}\right)\mathfrak{N}^{abc_{1} \cdots c_{n}ef}l_{c_{1}} \cdots l_{c_{n}}~,
\end{align}
the last equation implies that if $\delta g_{ef}$ belongs to the kernel of
$n_{a}n_{b}\mathfrak{N}^{abc_{1} \cdots c_{n}ef}l_{c_{1}}...l_{c_{n}}$ then it
belong to the kernel of $n_{a}k_{b}\mathfrak{N}^{abc_{1}...c_{n}ef}l_{c_{1}} \cdots l_{c_{n}}$.\footnote{This idea goes back to the fact that the momentum constraints guarantee the preservation of the Hamiltonian constraint.}
Thus, the solutions of the constraint characteristic equation~\eqref{eq:char_const} are the same solutions of
\begin{align}
    S_{i}^{ef}\delta g_{ef}\equiv\left[
    \begin{array}
    [c]{c}%
    n_{a}n_{b}\mathfrak{N}^{abc_{1} \cdots c_{n}def}l_{c_{1}} \cdots l_{c_{n}}\\
    e_{1(a}n_{b)}\mathfrak{N}^{abc_{1} \cdots c_{n}def}l_{c_{1}} \cdots l_{c_{n}}\\
    e_{2(a}n_{b)}\mathfrak{N}^{abc_{1} \cdots c_{n}def}l_{c_{1}} \cdots l_{c_{n}}%
    \end{array}
    \right]\delta g_{ef}=0~. \label{char_con_eq_1}
\end{align}
Since $S_{i}^{ef}$ is a $3\times10$ matrix, there
are at least seven linearly independent solutions~$\delta g_{ef}$ for any value of
$\lambda$. More solutions may exist for some specific values of $\lambda$ or, for instance, when a particular theory has any extra symmetries. We know that four solutions of \eqref{char_con_eq_1} are the gauge modes; they are $l_{(e}\zeta_{f)}$ for any four linearly independent $\zeta_{f}$. The direct sum of the previous four gauge modes and the following six elements
\begin{align}
    \label{eq:basis_P1}
B=\left\{
    \begin{array}
    [c]{c}%
    e_{1a}e_{1b}-e_{2a}e_{2b},~2e_{1(a}e_{2b)},\\
    e_{1a}e_{1b}+e_{2a}e_{2b},~n_{a}n_{b},\\
    2n_{(a}e_{1b)},~2n_{(a}e_{2b)}%
    \end{array}
    \right\}~,
\end{align}
furnishes a basis of the $\left(0,2\right)$-tensors $\text{Sym}\left(T_{p}^{\ast}M\times T_{p}^{\ast}M\right)$. Then, the other three solutions of equation~\eqref{char_con_eq_1}, denoted by $\left( \delta g_{1}\right)  _{ef}$, $\left(  \delta
g_{2}\right)  _{ef}$, and $\left(  \delta g_{C}\right)  _{ef}$, can be expanded as a linear combination of the elements of $B$ which span a six-dimensional vector subspace. These solutions indicate the physical sector of the principal symbol, as we show in the following examples.
%
%
%
%-------------------------------------------------------------------------------------------------
%-------------------------------------------------------------------------------------------------

\PRLsection{General Relativity}
We start by considering GR in vacuum. We have
\begin{align}
    G^{ab} = R^{ab}-\frac{1}{2}g^{ab}R = 0~,
\end{align}
and the principal symbol of the Einstein tensor is~\cite{Papallo:2017ddx}
\begin{align}
    \label{eq:GR_PS}
    \mathfrak{N}^{abc_{1}c_{2}ef}l_{c_{1}}l_{c_{2}}\equiv\left(-\frac{1}{2}G^{abef}l^{2}+G^{abcq}l_{c}G_{~~~~q}^{efd}l_{d}\right)~,
\end{align}
with
\begin{align}
    G^{abcd}=\frac{1}{2}\left(  g^{bc}g^{ad}+g^{ac}g^{db}-g^{ab}g^{cd}\right)~.
\end{align}
We solve equation~\eqref{char_con_eq_1}  and  find
\begin{align}
    \delta g_{1ef}  &  =\left(  e_{1e}e_{1f}-e_{2e}e_{2f}\right)~, \label{eq:g1}\\
    \delta g_{2ef}  &  =2e_{1(e}e_{2f)}~, \label{eq:g2}\\
    \delta g_{Cef} & =n_{e}n_{f}~. \label{eq:gC}
\end{align}
Then, the second-order pencil in the basis~\eqref{eq:basis_P1} is $ \left.  \mathfrak{N}^{abc_{1}c_{2}ef}l_{c_{1}}l_{c_{2}}\right\vert _{BB}~\equiv~ B_{ab}\mathfrak{N}^{abc_{1}c_{2}ef}l_{c_{1}}l_{c_{2}}B_{ef}$
which evaluates to
\begin{align}
    \label{eq:GR_PP_in_B}
    \left[
    \begin{array}
    [c]{cc|cccc}%
    \lambda^{2}-1 & 0 & 0 & 0 & 0 & 0\\
    0 & \lambda^{2}-1 & 0 & 0 & 0 & 0\\
    \hline
    0 & 0 & 1-\lambda^{2} & -1 & 0 & 0\\
    0 & 0 & -1 & 0 & 0 & 0\\
    0 & 0 & 0 & 0 & 1 & 0\\
    0 & 0 & 0 & 0 & 0 & 1
    \end{array}
    \right]~.
\end{align}
Notice that the solutions~\eqref{eq:g1} and~\eqref{eq:g2} are the first two elements of the basis~\eqref{eq:basis_P1}. This choice produces the final block structure in~\eqref{eq:GR_PP_in_B}, isolating the $2~\times~2$ upper-left block, i.e., the physical block, 
\begin{align}
    \label{eq:GR_PB}
    P_{\text{GR}} = \left(\lambda^2 - 1 \right)~\mathbb{I}_2~,
\end{align}
where $\mathbb{I}_2$ is the $2 \times 2$ identity matrix. This block contains the spin-$2$ dynamical physical information that is unchanged by gauge fixing or constraint addition. We also observe that the physical block of GR is wave-like since $\lambda^2-1$ is the Fourier space representation of the d'Alembert (box $\Box$) operator. Solving $\det P=0$, shows that the characteristic velocities of the physical sector are $\lambda=\pm1$ with algebraic multiplicity (or degeneracy) $q_{\pm1}=2$, and geometric multiplicity of $s_{\pm1}=2$. Ref.~\cite{Abalos:2026kim} showed that having equal multiplicities implies strong hyperbolicity; thus, the physical sector of GR is strongly hyperbolic. Hence, after a suitable gauge fixing, the full set of equations remains strongly hyperbolic. If the physical sector does not satisfy $s_{\lambda_i}~=~q_{\lambda_i}$, then there is no gauge fixing to render the system strongly hyperbolic. Nonetheless, the degeneracy might increase after fixing the gauge or due to the extra characteristics associated with the constraints, and the condition $s_{\lambda_i}~=~q_{\lambda_i}$ may fail. For instance, the ADM formulation is only weakly hyperbolic~\cite{Reula:1998ty,PhysRevD.70.044029,Frittelli:2000uj,bona2009elements,Shinkai:2008yb}, despite the physical sector of GR being healthy~\eqref{eq:GR_PB}. If the multiplicities are not equal, then the system is weakly hyperbolic. The generalization of these results to higher-order PDEs is straightforward and left for a follow-up paper.

%
%
%
%-------------------------------------------------------------------------------------------------
%-------------------------------------------------------------------------------------------------
\PRLsection{Higher-order EFTs of gravity}
Higher-order derivative terms appear naturally in the context of EFTs. Consider, for instance, the EFT as a derivative expansion in the metric. The leading-order corrections to GR are all the four-derivative terms, and (up to boundary terms and redundant interactions) are given by quadratic gravity~\cite{PhysRevD.28.1876,PhysRevD.32.379,PhysRevD.16.953,1978GReGr...9..353S,Noakes:1983xd,Held:2023aap, Held:2025ckb}. The Lagrangian is
\begin{align}
    \label{eq:L_quad_grav}
    \mathcal{L}_{\text{quad.}} = R + \alpha_0 R^{ab}R_{ab} - \beta_0 R^{2}~,
\end{align}
where $\alpha_0$ and $\beta_0$ are coupling constants. The equations of motion are
\begin{align}
    \label{eq:QG}
   0 &=  G_{ab} -2\beta_0 \left(  \nabla_{a}\nabla_{b}G-g_{ab}\Box G\right) \nonumber \\
   &+\alpha_0 \left(  \Box G_{ab}-g_{ab}\Box G-2\nabla^{c}\nabla_{(a}G_{b)c}+\nabla_{a}\nabla_{b}G\right)~.
\end{align}
Although the high-frequency limit lies outside the EFT regime, since it probes wavelengths comparable to or shorter than the cutoff scale, the well-posedness analysis remains relevant when solving the equations numerically. Hence, similar to GR, we solve the constraint characteristic equation~\eqref{eq:char_const}, and obtain the same solutions for $\delta g_{1ef}$ and $\delta g_{2ef}$; however, for $\delta g_{Cef}$ we find 
\begin{align}
    \delta g_{Cef} & =-\left(\alpha_0 +2 \beta_0\right) \left(  e_{1e}e_{1f}+e_{2e}e_{2f}\right) \nonumber \\
    &+ \left(\lambda ^2-1\right) (\alpha_0 +4 \beta_0 )n_{e}n_{f}~.
\end{align}
Therefore, to avoid a basis that is $\lambda$ dependent, we use the basis~\eqref{eq:basis_P1}, obtaining a similar block structure, with zeros in the off-diagonal blocks. This also occurs in the six-derivative and general higher-order EFTs discussed below. Where now the
 $2 \times 2$ physical block of $\left(\mathfrak{N}_{\text{quad.}}\left(l\right) \cdot \delta g\right)\vert _{BB}$ becomes
\begin{align}
    \label{eq:quad_P}
    P_{\text{quad.}} = -\alpha_{0}\left(\lambda^2 - 1 \right)^2~\mathbb{I}_2~.
\end{align}
Further, we consider the six-derivative gravity theories. At this order, Refs.~\cite{Carminati:1991ddy, Fulling:1992vm}, provide a basis of independent terms. After integrations by parts and the use of geometric identities, it can be reduced to
$\{R \Box R,~R^{ab} \Box R_{ab},~R^3,~\cdots \}$, where ellipses include terms with only contractions of curvature tensors. 
Among these, only \(R\Box R\) and \(R^{ab}\Box R_{ab}\) contribute to the highest-derivative principal part,\footnote{In vacuum it is often argued that such terms can be removed by perturbative field redefinitions~\cite{Burgess:2007pt,Burgess:2003jk}. We do not perform these redefinitions here, since, while they leave the physical content unchanged, they may alter the PDE structure and hence the well-posedness properties of the resulting formulation.} so the remaining algebraic curvature terms can be discarded for the hyperbolicity analysis. Therefore, the highest derivative terms appearing in the equations are due to these terms, and we consider
\begin{align}
\mathcal{L}_{\text{cubic}} = \alpha_1 R^{ab} \Box R_{ab} - \beta_1 R \Box R~.
\end{align}
Similar to the quadratic gravity case, we find for $\left(\mathfrak{N}_{\text{cubic}}\left(l\right) \cdot \delta g\right)\vert _{BB}$
\begin{align}
    \label{eq:cubic_P}
    P_{\text{cubic}} = -\alpha_{1}\left(\lambda^2 - 1 \right)^3~\mathbb{I}_2~,
\end{align}
Finally, for \emph{general higher-order EFTs of gravity}, the terms that produce the principal part are of the form  
\begin{align}
    \label{eq:L_higher_der}
    \mathcal{L}_{\text{HO}} = \sum_{k=0}^n\left(\alpha_{k}R^{ab}\Box^k R_{ab} - \beta_{k}R\Box^k R\right)~,
\end{align}
where $\alpha_k$ and $\beta_k$ are some coupling constants. From an EFT perspective, we emphasize that~\eqref{eq:L_higher_der} is the most general highest derivative terms. Additionally, these terms will only lead to equations of motion with characteristic velocities independent of derivatives of the metric~\cite{Figueras:2024bba}.
The resulting equations of motion are
\begin{align}
    \label{eq:higher_der}
    E^{ab}_{k} &\equiv \alpha_k\left(\Box^{k+1}G^{ab} - g^{ab} \Box^{k+1}G+ g^{ab}\nabla^{c}\nabla^{d}\Box^{k}G_{cd}
    \right.  \nonumber \\
    &\left. - \nabla^{c}\nabla^{a}\Box^k G^{b}{}_{c}
    - \nabla^{c}\nabla^{b}\Box^{k}G^{a}{}_{c}+\nabla^{a}\nabla^{b}\Box^{k}G\right) \nonumber \\
    &-2 \beta_{k}\left( \nabla^{a}\nabla^{b}\Box^{k} G - 2 g^{ab} \Box^{k+1} G\right) + \dots = 0~,
\end{align}
where ellipses denote lower-order terms. Therefore, for $\left(\mathfrak{N}_{\text{HO}}\left(l\right) \cdot \delta g\right)\vert _{BB}$, we have
\begin{align}
    \label{eq:HO_P}
    P_{\text{HO}} =  -\alpha_n(\lambda^2 - 1)^{2+n}~\mathbb{I}_2~.
\end{align}
Thus, the characteristic velocities are $\pm1$ with algebraic multiplicity (or degeneracy) $2 \left(2+n\right)$. However, the geometric multiplicity is $2$; consequently, the system is weakly hyperbolic, as a generalization of the results in~\cite{Abalos:2026kim}. Moreover, the physical block as in equations~\eqref{eq:quad_P},~\eqref{eq:cubic_P}, and~\eqref{eq:HO_P} have the structure of the scalar equation $\Box^{m}\phi + \text{lower order terms}=0,~ m>1$, which is weakly-hyperbolic, implying that the system is ill-posed for arbitrary lower terms. Nonetheless, Ref.~\cite{Figueras:2024bba} studied systems of the form~\eqref{eq:higher_der} and formulated the equations as a system of nonlinear wave equations for a specific set of variables. This shows that the system is well-posed in the Sobolev norm associated with those variables, but not necessarily in the Sobolev norm of the metric. 
%
%
%-------------------------------------------------------------------------------------------------
%-------------------------------------------------------------------------------------------------

\PRLsection{Discussion}
In this work, we studied the hyperbolicity of higher-derivative gravity EFTs. We solved the characteristic equation of these theories and disentangled the physical information from the gauge choice and constraint propagation. This was possible by employing the results of~\cite{Abalos:2026kim}, regarding weak and strong hyperbolicity without explicitly performing order reduction, and generalizing the various symmetries of the principal symbol~\cite{Reall:2021voz}. 

Higher-derivative EFTs of gravity, without performing field redefinitions, have a remarkable property: the principal symbol does not depend on derivatives of the metric, and the physical characteristic velocities are those associated with the physical metric. This property enables us to solve the characteristic equation without fixing the gauge or making any assumptions about the background. When theories violate this assumption, one must make additional assumptions about the background or how the derivatives behave. In general, if the principal symbol depends on derivatives of the metric, the characteristic velocities may cross, and as a result, the solutions develop shocks. For these reasons, we do not consider such theories here.

As an application of our approach, we demonstrated that in the $m$-th Sobolev energy norm $E(t)_{\mathcal{H}^m}$ (see~\eqref{eq:norm} and the trivial higher-order generalization thereof), due to the physical modes, higher-order EFTs of gravity are weakly hyperbolic. That is, for \textit{arbitrary lower-order terms}, these theories are ill-posed, i.e., there is no energy estimate as in~\eqref{eq:estimate} for the metric and all its derivatives, regardless of the gauge (an ill-advised gauge choice can only make things worse). Interestingly, this shows that, generally, in EFTs of gravity, the best one can achieve is a weakly hyperbolic system in $E(t)_{\mathcal{H}^m}$. However, Ref.~\cite{Figueras:2024bba} showed that a specific reduction yields a system of nonlinear wave equations that are well-posed in a \textit{different} norm, i.e., with a different $E(t)$, which requires additional regularity assumptions on the initial data. From a numerical relativity perspective, it is always preferable to have strongly hyperbolic systems. The adapted energy estimates may still allow convergence in a weaker norm, provided that both the data and the discretization possess enough additional regularity~\cite{Figueras:2025wtx}. However, the loss of derivatives makes convergence sensitive to high-frequency errors and may reduce the observed convergence order~\cite{Giannakopoulos:2023zzm, Gundlach:2024xmo}. Moreover, related EFT prescriptions aimed at controlling short-wavelength modes can introduce additional high-frequency scales, thereby improving UV control at the price of making the resulting evolution equations stiff~\cite{Besharat:2026rba}.

We leave a more thorough discussion of what constitutes an ``adapted'' norm, and the associated energy estimates for future work. On the practical side, exploring, if any, the numerical shortcomings remains pending and requires further investigation. 
%-------------------------------------------------------------------------------------------------
%-------------------------------------------------------------------------------------------------

\PRLsection{Acknowledgments}
We thank A. Kovács, A.Held, C. Palenzuela, D. Hilditch, H. Shum,  L. Aresté Saló, L. Lehner and P. Figueras for helpful discussions and comments. FT is supported by the INFN Postdoctoral Research Agreement No. 27076. FA acknowledges partial support from project PID2022-138963NB-I00, funded by the Spanish Ministry of Science, Innovation and Universities (MCIN/AEI/10.13039/501100011033). MB acknowledges partial support from the STFC Consolidated Grant nos. ST/Z000424/1 and UKRI2492.
\bibliography{biblio.bib}
\onecolumngrid
\appendix
\section{Principal symbol and symmetries}
\label{sec:syms}
In this section, we review definitions and results about the principal symbol and its symmetries. First, we reproduce some results obtained in~\cite{Reall:2021voz}, and then extend them to higher-derivative theories. These symmetries will prove useful when attempting to disentangle the physical modes from the constraints and the gauge choice. We will mainly focus on a general theory of only the metric $g_{ab}$
\begin{align}
    \label{eq:general_L_ap}
     S = \int \mathrm{d}^d x \sqrt{\vert g\vert } \mathcal{L}(g)~,
\end{align}
with its equations of motion given by
\begin{align}
        E^{ab} \equiv \frac{\delta S}{\delta g_{ab}} = \mathfrak{N}^{abc_{1} \cdots c_{n}ef}\partial_{c_{1}} \cdots \partial
    _{c_{n}}g_{ef}+ \cdots=0~. %
\end{align}
It follows directly from the preceding equation that 
\begin{align}
\mathfrak{N}^{abc_{1} \cdots c_{n}ef}    =\mathfrak{N}^{(ab)c_{1} \cdots c_{n}
ef} \nonumber 
=\mathfrak{N}^{abc_{1} \cdots c_{n}(ef)}.
\end{align}
\vspace{-2.em}
\begin{align}
\mathfrak{N}^{efc_{1} \cdots c_{i} \cdots c_{j} \cdots c_{n}ab}  =\mathfrak{N}%
^{efc_{1} \cdots c_{j} \cdots c_{i} \cdots c_{n}ab} \ \ \ \forall~i,~j \in \{1,~\cdots ,~n\}~.
\end{align}
Moreover, as shown below, the principal symbol $\mathfrak{N}^{abc_{1} \cdots c_{n}ef}$ satisfies the additional identities
\begin{align}
\mathfrak{N}^{abc_{1} \cdots c_{n}ef}  
&=\mathfrak{N}^{efc_{1} \cdots c_{n}ab}, \\
0  &  =\mathfrak{N}^{a(bc_{1} \cdots c_{n})ef},\\
0  &  =\mathfrak{N}^{ab(c_{1} \cdots c_{n}e)f}~.
\end{align}

\subsection{Second-order equations of motion}
Let us, as a start, consider gravity theories with second-order equations. Then the symbol is given by 
\begin{align}
    \label{eq:PS_def_0}
\mathfrak{N}^{abc_1c_2ef}l_{c_1}l_{c_2} = \frac{\partial E^{ab}}{\partial\left(\partial_{c_1}\partial_{c_2}g_{ef}\right)}l_{c_1}l_{c_2}~,
\end{align}
it is clear from the definition that the symbol has the following symmetries
\begin{align}
    \mathfrak{N}^{(ab)c_1c_2ef} =\mathfrak{N}^{ab(c_1c_2)ef} = \mathfrak{N}^{abc_1c_2(ef)}~.
\end{align}
Furthermore, the symbol possesses additional symmetries arising from the action principle and diffeomorphism invariance~\cite{Reall:2021voz}. In what follows, we will first review the arguments behind the symmetries and then generalize them to higher-derivative theories. Given that the equations arise from an action principle, a first variation of the action would give (dropping boundary terms)
\begin{align}
    \label{eq:var1}
    \delta_1S = \int \mathrm{d}^dx \sqrt{\vert g\vert}\left(E^{ab}\delta_1g_{ab}\right)~,
\end{align}
performing a second variation, we obtain
\begin{align}
    \label{eq:double_var}
    \delta_2\delta_1S = \int \mathrm{d}^dx \sqrt{\vert g\vert}\left(E^{ab}\delta_2\delta_1g_{ab} + \left(\mathfrak{N}^{abc_1c_2ef}\partial_{c_1}\partial_{c_2}\delta_2g_{ef}+\dots\right)\delta_1g_{ab}\right)~,
\end{align}
where ellipses denote terms with fewer than two derivatives acting on a variation of the metric. As we are ignoring such terms, we can replace the partial derivatives with covariant ones and integrate by parts to get 
\begin{align}
\nabla_{c_1}\left(
\mathfrak{N}^{abc_1c_2ef}
\nabla_{c_2}\delta_2 g_{ef}\,
\delta_1 g_{ab}
\right)
&=
\mathfrak{N}^{abc_1c_2ef}
\nabla_{c_2}\delta_2 g_{ef}\,
\nabla_{c_1}\delta_1 g_{ab} 
\nonumber \\
&\quad
+
\mathfrak{N}^{abc_1c_2ef}
\nabla_{c_1}\nabla_{c_2}\delta_2 g_{ef}\,
\delta_1 g_{ab}
\nonumber 
+
\delta_1 g_{ab}\,
\nabla_{c_1}\mathfrak{N}^{abc_1c_2ef}
\nabla_{c_2}\delta_2 g_{ef}~.
\end{align}
To deduce the symmetries of the symbol arising from the action principle, we need not keep track of terms with no derivatives acting on the metric variations~\cite{Reall:2021voz}; thus, we have
\begin{align}
      \mathfrak{N}^{abc_1c_2ef}
\nabla_{c_1}\nabla_{c_2}\delta_2 g_{ef}\,
\delta_1 g_{ab}  \sim - \mathfrak{N}^{abc_1c_2ef}
\nabla_{c_2}\delta_2 g_{ef}\,
\nabla_{c_1}\delta_1 g_{ab}   + \dots~.
\end{align}
To obtain the symmetries arising from the action principle, we antisymmetrize the two variations and evaluate them on the background. Now, due to the (trivial) symmetries of the symbol $\mathfrak{N}^{abc_1c_2ef} = \mathfrak{N}^{abc_2c_1ef}$, we can write the antisymmetrization of the second term in~\eqref{eq:double_var} (the $\mathfrak{N}$ term) as
\begin{align}
    \mathfrak{N}^{abc_1c_2ef}
\nabla_{c_2}\delta_2 g_{ef}\,
\nabla_{c_1}\delta_1 g_{ab} - \mathfrak{N}^{abc_1c_2ef}
\nabla_{c_2}\delta_1 g_{ef}\,
\nabla_{c_1}\delta_2 g_{ab}  
    =(\mathfrak{N}^{abc_1c_2ef}-\mathfrak{N}^{efc_1c_2ab}) \nabla_{c_1}\delta_1 g_{ab} \nabla_{c_2}\delta_2 g_{ef},
\end{align}
and the term $E^{ab}\delta_2\delta_1g_{ab}$ drops out when antisymmetrized, hence,
\begin{align}
    0 = \int \mathrm{d}^d x \sqrt{\vert g\vert} \left[(\mathfrak{N}^{abc_1c_2ef}-\mathfrak{N}^{efc_1c_2ab}) \nabla_{c_1}\delta_1 g_{ab} \nabla_{c_2}\delta_2 g_{ef}+\dots\right]~,
\end{align}
which must hold for any compactly supported variations. Therefore, the coefficients of the terms quadratic in the first derivatives of the variations must vanish~\cite{Reall:2021voz}, and we have 
\begin{align}
    \label{eq:sym_action}
    \mathfrak{N}^{abc_1c_2ef} = \mathfrak{N}^{efc_1c_2ab}~,
\end{align}
that is, the symbol is symmetric in its first two pairs of indices.

Next we consider the symmetries arising due to diffeomorphism invariance. Consider a diffeomorphism due to a compactly supported vector field $X^{a}$, then the variation in the metric is
\begin{align}
    \delta g_{ab} = 2 \nabla_{(a}X_{b)}~.
\end{align}
If we take $\delta_1 g_{ab}$ in~\eqref{eq:var1} to be such a diffeomorphism, and integrate by parts we find a generalized Bianchi identity
\begin{align}
    \nabla_{b}E^{ab} = 0~, \label{eq_Bianchi_identity}
\end{align}
expanding this equation, we have
\begin{align}
    \mathfrak{N}^{abc_1c_2ef}\partial_{bc_1c_2}g_{ef} + \cdots = 0~,
\end{align}
where ellipses denote lower-order terms. Since these equations must hold for arbitrary configurations, the coefficient of the third-order derivatives must vanish, leading to
\begin{align}
    \mathfrak{N}^{a(bc_1c_2)ef} = 0~.
\end{align}
Using the symmetry~\eqref{eq:sym_action}, we have
\begin{align}
    \mathfrak{N}^{ab(c_1c_2e)f} = 0~.
\end{align}
\subsection{Higher-order equations}
For higher-derivative theories, we show that the same symmetries must hold for fourth-order equations; however, the discussion below is easily generalizable to any order in derivatives. The symbol of higher-derivative theories is
\begin{align}
    \label{eq:PS_def_1}  \mathfrak{N}^{abc_1\dots c_nef}l_{c_1}\dots l_{c_n} = \frac{\partial E^{ab}}{\partial\left(\partial_{c_1}\dots\partial_{c_n}g_{ef}\right)}l_{c_1}\dots l_{c_n}~.
\end{align}
We apply the procedure outlined in the previous section to the fourth-order theory and obtain the following double variation of the action  
\begin{align}
    \label{eq:double_var_2}
    \delta_2\delta_1S = \int \mathrm{d}^dx \sqrt{\vert g\vert}\left(E^{ab}\delta_2\delta_1g_{ab} + \left(\mathfrak{N}^{abc_1c_2c_3c_4ef}\partial_{c_1c_2c_3c_4}\delta_2g_{ef}+\dots\right)\delta_1g_{ab}\right)~,
\end{align}
where now ellipses denote terms with fewer than four derivatives acting on a variation of the metric. As before, and for the same reasons, we exchange the $\partial$'s with $\nabla$'s, and integrate twice by parts, to have
\begin{align}
    \mathfrak{N}^{abc_1c_2c_3c_4ef}\nabla_{c_1c_2c_3c_4}\delta_1g_{ab}\delta_2g_{ef} \sim  \mathfrak{N}^{abc_1c_2c_3c_4ef}\nabla_{c_3c_4}\delta_1g_{ab} \nabla_{c_1c_2}\delta_2g_{ef}
    + \dots~.
\end{align}
Anti-symmetrizing the double variation will give 
\begin{align}
\mathfrak{N}^{abc_1c_2c_3c_4ef}\nabla_{c_3c_4}\delta_1g_{ab} \nabla_{c_1c_2}\delta_2g_{ef}-\mathfrak{N}^{abc_1c_2c_3c_4ef}\nabla_{c_3c_4}\delta_2g_{ab} \nabla_{c_1c_2}\delta_1g_{ef} \\    =\left(\mathfrak{N}^{abc_1c_2c_3c_4ef}-\mathfrak{N}^{efc_1c_2c_3c_4ab}\right)\nabla_{c_3c_4}\delta_1g_{ab} \nabla_{c_1c_2}\delta_2g_{ef}~,
\end{align}
therefore, we have 
\begin{align}
    \mathfrak{N}^{abc_1c_2c_3c_4ef}=\mathfrak{N}^{efc_1c_2c_3c_4ab}~,
\end{align}
and, as in the second-order equations case, the symbol is symmetric in its first two pairs of indices. The symmetry due to diffeomorphisms follows similarly to the second-order case, and we have 
\begin{align}
    \mathfrak{N}^{a(bc_1c_2c_3c_4)ef} = 0~,
\end{align}
combining both symmetries leads to 
\begin{align}
    \mathfrak{N}^{ab(c_1c_2c_3c_4e)f} = 0~.
\end{align}
These results are clearly generalizable to any order in the derivatives.
\end{document}